\documentclass[manuscript]{aastex}

\usepackage{graphicx}
\usepackage{color}
\begin{document}

\shorttitle{Spectrum of Relativistic and Subrelativistic Cosmic Rays...} \shortauthors{Dogiel et al.}

\title{Spectrum of Relativistic and Subrelativistic Cosmic Rays in the 100 pc Central Region }
\author{V. A. Dogiel$^{1,2,3,7}$, D. O. Chernyshov$^{1,2}$, A. M. Kiselev$^{1}$, M. Nobukawa$^{4,5}$, K. S. Cheng$^{2}$,
C. Y. Hui$^6$, C. M. Ko$^{3}$, K. K. Nobukawa$^{4}$, T. G. Tsuru$^{4}$}
\affil{$^1$I.E.Tamm Theoretical Physics Division of P.N.Lebedev
Institute of Physics, Leninskii pr. 53, 119991 Moscow, Russia}
%\email{dogiel@lpi.ru}
\affil{$^2$Department of Physics, University of Hong Kong,
Pokfulam Road, Hong Kong, China}
\affil{$^3$Institute of Astronomy, Department of Physics and Center for Complex Systems, National Central University,
Jhongli, Taiwan}
%\email{cmko@astro.ncu.edu.tw}
\affil{$^4$Department of Physics, Graduate School of Science, Kyoto University, Kitashirakawa-oiwake-cho, Sakyo-ku, Kyoto 606-8502, Japan}
\affil{$^5$The Hakubi Center for Advanced Research, Kyoto University, Yoshida-Ushinomiya-cho, Kyoto 606-8302, Japan}
\affil{$^6$Department of Astronomy and Space Science, Chungnam National University, Daejeon, Korea}
\affil{$^7$Moscow Institute of
Physics and Technology, 141700 Moscow Region, Dolgoprudnii,
Russia}

%%%%%%%%%%%%%%%%%%%%%%%%%%%%%%%%%%%%%%%%%%%%%%%%%%%%%%%%%%%%%%%%%%%%%%%%%%%%%%%%%%%%%%%%%%
%%%%%%%%%%%%%%%%%%%%%%%%%%%%%%%%%%%%%%%%%%%%%%%%%%%%%%%%%%%%%%%%%%%%%%%%%%%%%%%%%%%%%%%%%%
%%%%%%%%%%%%%%%%%%%%%%%%%%%%%%%%%%%%%%%%%%%%%%%%%%%%%%%%%%%%%%%%%%%%%%%%
\altaffiltext{1}{V.A.D. email: dogiel@lpi.ru}
\altaffiltext{3}{C.M.K. email: cmko@astro.ncu.edu.tw }

\begin{abstract}
From the rate of hydrogen ionization and the gamma ray flux,
we derived the spectrum of relativistic and subrelativistic cosmic rays (CRs) nearby and inside the molecular cloud Sgr B2 near the Galactic Center (GC).
We studied two cases of CR propagation in molecular clouds: free propagation
and scattering of particles by magnetic fluctuations excited by the neutral gas turbulence.
We showed that in the latter case CR propagation inside the cloud can be described as
diffusion with the coefficient $\sim 3\times 10^{27}$ cm$^2$ s$^{-1}$.
For the case of hydrogen ionization by subrelativistic protons,  we showed that their
spectrum outside the cloud is quite hard with the spectral index $\delta>-1$.
The energy density of subrelativistic protons ($>50$ eV cm$^{-3}$) is one
order of magnitude higher than that of relativistic CRs. These protons
generate the 6.4 keV emission from Sgr B2, which was about 30\% of the flux
observed by Suzaku in 2013.  Future observations for the period after 2013 may
discover the background flux generated by subrelativistic CRs in Sgr B2.
Alternatively hydrogen ionization of the molecular gas in Sgr B2 may be caused
by high energy electrons. We showed that the spectrum of electron
bremsstrahlung is harder than the observed continuum from Sgr B2, and
in principle this X-ray component provided by electrons could be seen from the
INTEGRAL data as a stationary high energy excess above the observed spectrum $E_x^{-2}$.
\end{abstract}

\date{\today}

\maketitle

\section{Introduction}\label{intro}

One of the important but still unresolved problems of astrophysics is spatial distribution of cosmic rays (CRs) in the Galaxy.
We still do not have reliable understanding on how CRs are distributed in different parts of the Galaxy and what is their spectrum there.
In principle, this can be estimated from the characteristics of nonthermal emission generated by CRs in a range of wavelengths.
The density of relativistic protons in the Galactic Disk (GD) can be derived from the observed intensity of gamma-rays,
which are believed to be produced by proton-proton ($pp$) collisions while the density of subrelativistic CRs can be
estimated from the emission of nuclear de-excitation lines and IR-absorption lines of ionized hydrogen \citep[see, e.g.,][]{ber90,ram79, oka05}.

Several attempts were undertaken to derive the distribution of relativistic CRs in GD with energies above 1 GeV from the gamma-ray data.
The distribution of CRs, derived from the COS-B \citep[][]{bhat86,bloe86,strong88}, EGRET \citep[see, e.g.,][]{strong96}
and Fermi-LAT \citep[][]{abdo10,acker11,tib13} measurements, showed that the density of CRs in the Galactic Center (GC) was higher than
near Earth but their spatial distribution in the GD was much flatter than the distribution of their potential Galactic sources:
supernova remnants (SNRs) \citep[see][]{case98,green12} or pulsars \citep[see, e.g.,][]{lor04}.

A natural explanation could be an effective spatial mixing of CRs by particle scattering on magnetic fluctuations in the Galaxy if
the GD is surrounded by a giant halo, in which CRs spend a significant part of their lifetime before escaping from the
Galaxy \citep[see][]{ber90}. However, numerical calculations showed \citep[see, e.g.,][]{dog88,bloemen}, even in the most favorable
case of a very extended halo, that this scattering (described as diffusion propagation) was unable to remove the signature of the
source distribution.
The CR distribution derived from the diffusion model was steeper than the distribution inferred from gamma-ray observations,
although it was flatter than the source distribution.
One of the explanations was suggested by \citet{strong04} who showed that the problem can be solved by a
variation in the $W_{CO}$-to-$N(H_2)$ (metalicity) scaling factor.
Another explanation was suggested by \citet{breit02} who assumed that CRs left the disk faster from regions of a higher
concentration of SNRs which will smoothen the CR distribution in the GD in comparison with that of their sources.

However, estimates of the CR density, derived from the gamma-ray data, are model dependent, because a simplified model
of the gas distribution in the GD is used in calculations. Besides, the adopted distribution of CR sources in the
GD cannot be accepted as absolutely reliable because of the obscurity of dust and so on.

Another possibility to define the density of CRs at different galactocentric radii
(and even at different altitudes above the GD)
is the analysis of gamma-ray emission
from molecular clouds whose total mass is known, see \citet{digel,hui,yang1,yang2,tib15} for the GeV energy range and
\citet{ahar06} for the TeV energy range.
A special case is the circumnuclear disk of molecular gas in the Galactic plane whose position may coincide with
the source J1745.6-2858 from the second and third Fermi LAT source catalog \citep[see][]{nolan12,abdo15}.
This gamma-ray emission can be provided by CRs generated in the course of present or past activity of Sgr A*
\citep[see][]{masha11,chern14,malysh15}.
However, analysis of the radiation from the source J1745.6-2858 is beyond the scope of the present work.
We suppose to analyse this emission elsewhere.
We note that \citet{eldik15} provided an excellent review on the origin of gamma ray emission from the GC.

Recent analysis of gamma-ray emission from local molecular cloud near Earth and the cloud Sgr B2 in the GC provided by
\citet{yang1,yang2} showed that the density of CRs was almost the same in these regions although
the distance between them is about 8 kpc.
However, these estimations of CR density inside the clouds depend on how freely can energetic particles penetrate into molecular clouds.
This question is one of the goals of this article.

The spectrum of CRs in the energy range below 1 GeV can also be estimated from the gamma-ray data.
\citet{dermer13} derived the spectrum  of proton component of CRs in the local interstellar gas
from the mid-latitude measurements of the Galactic gamma-ray emission \citep[see][]{abdo09}.
These investigations showed that there was a flattening in the proton spectrum below several GeV.
Similar conclusion was obtained by \citet{neron12} from investigation of gamma-ray emission generated in local molecular clouds.
However, due to the energy threshold of $pp$ reaction, the gamma-ray data that is useful for deriving the spectrum of CR protons are
those with energies above hundred MeV.

\citet{bouchet11} derived the spectrum of Galactic electrons in the range between 1 and 5 GeV
from the intensity of diffuse hard X-ray emission measured by INTEGRAL.
They showed that a significant fraction of the Ridge emission in the range above 50 keV is produced
by the inverse Compton scattering of these electrons on background photons.

Almost forty years ago \citet{ram79} suggested to measure a flux of nuclear de-excitation lines with $E_\gamma\sim 0.1-10$ MeV generated by
subrelativistic CRs. In order to estimate their density in the range between several MeV and hundred MeV.
However, the sensitivity of gamma-ray telescopes was not (and still is not) high enough to detect these lines in the diffuse gamma-ray spectrum
\citep[see e.g.][]{dog09b,ben13}.

Another very interesting results about the parameters of low energy cosmic rays in the GC region was obtained by \citet{nobu15}.
They found an excess of the 6.4 keV line emission in the near east of the GC.
They concluded that this excess is due to iron atom bombardment by protons with energies 0.1 - 1000 MeV.
The estimated energy density of these protons is about 80 eV cm$^{-3}$ which is almost two orders of magnitude higher
than the energy density of relativistic CRs near Earth.

In principle, the spectrum of low energy CRs in the range below 100 MeV can also be derived from the ionization rate of hydrogen
in the interstellar medium. The point is that subrelativistic electrons and protons ionize hydrogen molecules effectively.
The ionization rate can be derived from the $H_3^+$ IR absorption lines
\citep[see][see also the review of \citet{dalg06}]{oka05,indri12} that allows one to estimate the
density of subrelativistic CRs inside molecular clouds and in the intercloud medium \citep[see e.g.][]{dog02,dog13,dog14}.

The average ionization rate in the 100 pc central region of the Galaxy, measured by \citet{oka05}, is about
$(3-5)\times 10^{-15}$ s$^{-1}$, while this value for the cloud Sgr B2 is depleted and equals $4\times 10^{-16}$ s$^{-1}$ \citep{vandertak}.
This means that penetration of subrelativistic CRs into Sgr B2 is not complete, and the average density of these CRs in the intracloud region
is higher than inside Sgr B2. Similar effect is observed for the clouds outside the GC where the ionization rate is about
$3.5\times 10^{-16}$ s$^{-1}$ in the diffuse clouds, while this value is $3\times 10^{-17}$ s$^{-1}$ in dense molecular clouds \citep{indri12}.

The spectrum of subrelativistic CRs inside molecular clouds depends on processes of particle propagation there.
Several models are used in order to estimate CR density in the clouds. One of them is extrapolation of the model of CR
propagation in the interstellar medium when charged particles are scattered by resonant MHD waves
\citep[see, e.g.,][]{dog02,gabici,protheroe}. Alternatively, \citet{padovani} derived the spectrum of subrelativistic CRs in
the local interstellar medium from the rate of hydrogen ionization in the local molecular clouds assuming that they propagate freely
(without scattering) inside the clouds although the effect of large scale magnetic field may be essential in this case \citep[see][]{padovani11}.
We discussed processes of CR penetration into the clouds and show that this process may differ strongly than supposed in previous publications.
We try to estimate also the spectrum of CRs in the GC from the observed gamma-ray flux and the ionization rate of hydrogen in Sgr B2.
Propagation of CRs in partially ionized medium is a rather complicated process, which includes CR ionization of the medium,
hydromagnetic waves generation and damping, density and magnetic structures of the medium.
We address the density structure and ionization state of Sgr B2 in Section 2.
Section 3 discusses excitation of hydromagnetic waves by CRs, wave damping by ion-neutral collisions and diffusion of CRs in the cloud,
and Section 4 considers the magnetic structure of the cloud.
Base on these, we calculate the CR spectrum for two models of the cloud in Section 5.
Section 6 gives some consequences of the model.
Section 7 discusses the possibility of ionization by electrons.
Finally, in Section 8 we provide a summary.

\section{Hydrogen parameters of Sgr B2 and energy losses of CRs}\label{gas}

Parameters of CR propagation  depends on the medium parameters.
The strength of magnetic field in Sgr B2 was estimated from the
Zeeman splitting and was about $0.5$ mG \citep[see, e.g.,][]{crutcher}.
The cloud Sgr B2 is a very massive complex.
\citet{ahar06} estimated its total mass as $(6-15)\times 10^6$ M$_\odot$ for a $0.5^\circ\times 0.5^\circ$ ($75\times 75$ pc$^2$)
region surrounding Sgr B2. The gas distribution in the cloud Sgr B2 is highly uncertain.

\citet{lis89,lis90,lis91} and \citet{gold90} derived a two-component density  distribution: an envelope with a constant
$H_2$ density about 1800-3500 cm$^{-3}$ extending to the outer radius about 22.5 pc and a central core with the central density
$4-9\times 10^4$ cm$^{-3}$. According to their analysis the density distribution can be approximated as
\begin{equation}
n_{H_2}(r) =
 \left\{
\begin{array}{ll}
n_1+n_2\,, & {\rm if}\quad r\leq r_0\,, \\
n_1\left(\frac{r_0}{r}\right)^\alpha+n_2\,, & {\rm if}\quad R\geq r>r_0\,,
\end{array}\right.
\label{nH2a}
\end{equation}
with $r_0=1.25$ pc, $R=22.5$ pc, $n_1=5.5\times 10^4$ cm$^{-3}$, $n_2=2.2\times 10^3$ cm$^{-3}$, and $\alpha = 2$.
The $H_2$ column density at the center is, $N(H_2)=2.6\times 10^{24}$ cm$^{-2}$, and the total mass of Sgr B2 they took as
$6.3\times 10^6$ M$_\odot$.

\citet{protheroe} expressed the gas distribution in Sgr B2 in the form
\begin{equation}
n_{H_2}=\frac{M_{H_2}}{2m_H}\frac{1}{(\sqrt{2\pi}\sigma)^3}\exp(-r^2/2\sigma^2)\,,
\label{nH2}
\end{equation}
where $M_{H_2}$ is the total mass of Sgr B2 and $\sigma =2.75$ pc.
The density is about  $10^5$ cm$^{-3}$ at the center of the Sgr B2 complex, and decreases to 10 cm$^{-3}$ at a radius of $\sim 12$ pc,
which they considered to be its outer radius. The central column density is about $2.5\times 10^{24}$cm$^{-2}$ and the
estimated mass of Sgr B2 is about $M_{H_2}\simeq 2\times 10^6$ M$_\odot$ \citep[see][]{protheroe}.

Below we analyze processes of hydrogen ionization and gamma-ray emission for these two extreme distributions of hydrogen in Sgr B2.

CRs lose effectively their energy inside the dense molecular clouds.
In the non-relativistic energy range the rate of losses is determined by ionization
\citep[see][]{haya, ginz}
\begin{equation}
\left({{dE}\over{dt}}\right)_i=-{{2\pi
e^4n_H}\over{m_ec\beta(E)}}\ln\left({{m_e^2c^2W_{\rm max}} \over{4\pi
e^2\hbar^{2} n}}\right) \label{iot}\,,
\end{equation}
where $n_H$ is the density of background gas, $m_e$ is electron mass, $W_{\rm max}$ is the highest energy transmitted to an ambient electron,
and $\beta(E)=u/c$, where $\beta$ is the particle velocity $u$ in the unit of light speed $c$. The characteristic time of the losses is
\begin{equation}
\tau_i=\int\limits_E\frac{dE}{(dE/dt)_i}\,.
\label{tio}
\end{equation}

In the relativistic energy range protons lose their energy by collisions with the ambient gas ($pp$ collisions).
The characteristic time of the process is
\begin{equation}
\tau_{pp}=cn_H\sigma_{pp}\,,
\label{tpp}
\end{equation}
where the cross-section of $pp$-reaction can be found in e.g. \citet{kamae}.

One of the key point for the problem of CR penetration into dense molecular clouds  is how they propagate through the gas.
This problem is discussed in the next section.

\section{CR diffusion inside molecular clouds}\label{CRdiff}

In the standard model of CR propagation through the interstellar medium \citep[see, e.g.,][]{ber90} the spatial diffusion coefficient
can be presented in the form
\begin{equation}
D\sim \frac{u^2}{\nu(E)}\,,
\label{D_st1}
\end{equation}
where $\nu(E)$ is  the frequency of particle  scattering by resonant MHD-waves:
\begin{equation}
\nu\simeq\omega_H\frac{\delta H(k)^2}{H_0^2}\,.
\end{equation}
Here $\delta H(k)$ is the strength of a magnetic fluctuation with the wave number $k$, $H_0$ is the large scale magnetic field,
$\omega_H=eH_0/mc\gamma$, $\gamma$ and $m$ are the gamma-factor and the mass of a scattered particle, respectively.
The wave-particle interaction is resonant. A particle with the energy $E$ is scattered by waves whose wavelength
is about the particle Larmor radius, $\lambda=2\pi /k\sim r_L(E)$.
The amplitude of magnetic fluctuations, $\delta H$, is supposed to be much smaller than the strength of the large scale magnetic field
$H_0$ in the interstellar medium,
\begin{equation}
\delta H(k)\ll H_0\,.
\label{H_res}
\end{equation}
For the wave spectrum of fluctuations $W(k)\propto k^{-\kappa}$, where $\delta H(k)^2=kW(k)$,
the spatial diffusion coefficient $D$ of relativistic CRs is energy dependent as \citep[see][]{ber90}
\begin{equation}
D(E)\propto E^{2-\kappa}\,.
\label{D_st}
\end{equation}

This model of CR diffusion in the interstellar medium is often extrapolated to the case of CR propagation inside molecular clouds.
However, the process of particle propagation nearby and inside almost neutral and dense gas of molecular clouds is quite
different in comparison with the standard process of CR diffusion in the interstellar medium.
There are several specific mechanisms which determine this process:
\begin{enumerate}
\item  Dense molecular clouds absorb CRs with energies whose lifetime $\tau_{lt}(E)$
(determined by Equation~(\ref{tio}) or (\ref{tpp})) is smaller than the time of particle
propagation through the cloud, $\tau_{pr}$.
For ballistic propagation in the cloud $\tau_{pr}=R/u$ and for diffusion propagation $\tau_{pr}=R^2/D_c$ \citep[see e.g.][]{morfill82,dog85}.
Here $R$ is a cloud radius, $u$ is the particle velocity, and $D_c$ is the coefficient of spatial diffusion in the cloud.

    For energies determined by the condition, $\tau_{lt}(E)<\tau_{pr}(E)$, a flux of CRs to the cloud
    arises because of CR absorption. This CR flux generates MHD-waves with the increment \citep[see][]{kuls69}
    \begin{equation}
    \gamma({\bf k}, {\bf r})=\frac{\pi^2e^2v_A}{c^2}\int\limits_pd{\bf p}u(1-\mu^2)\delta\left(p|\mu k_\||-\frac{eH}{c}\right)
    \left(\frac{k_\|}{|{\bf k}|} \frac{\partial f}{\partial\mu}+\frac{v_A}{u}p\frac{\partial f}{\partial p}\right)\,.
    \end{equation}
    Here $H$ is the magnetic field strength near the cloud, $v_A=H/\sqrt{4\pi \rho_i}$ is the Alfv\`en velocity,
    $\rho_i$ is the density of ionized gas, $f({\bf p},{\bf r},\mu)$ is the CR particle distribution function,
    ${\bf p}$ is the particle momentum, $\mu$ is the particle pitch-angle, ${\bf k}$ is the wave number of
    a magnetic fluctuation excited by CRs, and $k_\|={\bf H}\cdot{\bf k}/|{\bf H}|$.
\item On the other hand, ion-neutral friction in the dense low ionised molecular clouds damps
MHD-waves with frequencies $\omega=kV_A<\mu_{in}$, where $V_A=H/\sqrt{4\pi\rho_i}$ is the Alfv\`enic velocity,
$\rho_i$ is the density of ionised component \citep[see e.g.][]{kuls69}, and the decrement of wave absorption is
    \begin{equation}
    \mu_{in}\sim \frac {m_n}{(m_i + m_n)}n_H\langle\sigma_{in}v_H\rangle\,.
    \end{equation}
Here $\sigma_{in}$ is the cross-section of ion-neutral collisions, $m_n$ is the mass of neutral particles,
$m_i$ is the mass of ionised particles, and $v_H$ is the thermal velocity of hydrogen. For the condition of molecular clouds
$\langle \sigma v \rangle = 2\cdot 10^{-9}$cm$^3$s$^{-1}$ \citep[see][]{pinto08}.

    \citet{dog85} provided detailed analysis of CR propagation nearby molecular clouds and showed that MHD waves
    are excited far away from the clouds, but they are completely damped inside the clouds. This means that CRs
    should freely propagate inside the clouds without scattering (but with energy loss). We notice also that similar analysis of CR propagation
    nearby the clouds and MHD-wave excitation there was provided latter by \citet{zweibel} and \citet{morlino15}.
\item However, as observations showed, the neutral gas in the clouds is highly turbulent \citep[see the review of][]{falgarone}.
This turbulence of neutral gas excites forced magnetic fluctuations through interaction with the ionized component \citep{dog87}.
These fluctuations prevent free particle propagation in the clouds. This process is discussed in the next section.
\end{enumerate}

\section{Magnetic field structure inside molecular clouds}\label{diffusion}

Detailed analysis of magnetic field structure in turbulent molecular clouds was provided by \citet{kiselev13}.
Below we present a brief review of their analysis and apply its results to the case of Sgr B2.
For an homogeneous and isotropic medium they considered the correlation function of the velocity field
$\langle v_i(\textbf{x},t)v_j(\textbf{x}+\textbf{r},\bar{t}) \rangle$ as delta-correlated in time,
\begin{equation}
\langle v_i(\textbf{x},t)v_j(\textbf{x}+\textbf{r},\bar{t}) \rangle=v_{ij}(\textbf{r}) \tau_{\rm c} \delta(t-\bar{t})\,,
\end{equation}
where $\tau_c$ is the characteristic time of  hydrodynamic turbulence of the neutral gas.

From the conditions of the tensor symmetry and for the incompressible liquid ($\nabla\cdot {\bf v}=0$)
the correlation tensor $v_{ij}(r)$ can be presented in the form
\begin{equation}
v_{ij}(r)= 2 V(r) \delta_{ij} + r\frac{dV(r)}{dr} (\delta_{ij} - \frac{r_i r_j}{r^2})\,,
\end{equation}
where $V(r)$ is an unknown function, which will be derived below from the observed spectrum of turbulence in molecular clouds.

In this medium the correlation tensor of magnetic fluctuations $b_{ij}$ similar to the velocity correlation tensor,
but the average is taken at the same time moments,
\begin{equation}
b_{ij}(r,t)=\langle b_i(\textbf{x},t)b_j(\textbf{x}+\textbf{r},t) \rangle= 2 Q
\delta_{ij} + r\frac{dQ}{dr} \left(\delta_{ij} - \frac{r_i r_j}{r^2} \right)\,,
\end{equation}
where the function $Q$ is a function of $r$ and $t$ in the general case.
The amplitude of magnetic fluctuations $b_0$ is defined as $b_0^2=\langle b_i(\textbf{x})b_i(\textbf{x}) \rangle$.

If there is a large scale magnetic field  $H_0$ in the medium, then the equation for the correlation function
$Q$ can be derived from the MHD equations \citep[see][]{kiselev13}
\begin{eqnarray}
& {\displaystyle \frac{1}{2\tau_{\rm c}} \frac{\partial Q(r)}{\partial t}}
{\displaystyle = \left[V(0) - V(r) + \frac{1}{\pi \rho_{\rm i} \mu_{\rm in} \tau_{\rm c}} \left( Q(0) + \frac{H_0^2}{6}\right) \right]
\left(\frac{d^2Q}{dr^2} +\frac{4}{r}\frac{dQ}{dr} \right)}  \nonumber \\
& {\displaystyle -\frac{dV}{dr}\frac{dQ}{dr} - \frac{1}{r}\left(4\frac{dV}{dr} + r\frac{d^2V}{dr^2}\right)
\left(Q+\frac{H_0^2}{6}\right)\,,}
\label{b_corr}
\end{eqnarray}
where $\rho_i$ is the density of ionised component of the gas.

The spectrum of turbulent velocities in molecular clouds was derived from measurements  of the CO and NH$_3$ line Doppler broadening.
The turbulence has a power-law Kolmogorov-like spectrum in a very broad range of scales from supersonic
\citep{lars} to subsonic \citep{my} regions \citep[see also the review of][]{falgarone}:
\begin{equation}
v(L)=1.1\ L^\alpha({\rm pc})\ {\rm km}\ {\rm s}^{-1}\quad {\rm where}\quad \alpha\simeq 0.3 - 0.5\,,
\label{ul}
\end{equation}
where $v$ is the velocity of turbulent motions and $L$ is its scale ($0.01<L<300$ pc).
Then from Equation~(\ref{ul}) the correlation function of velocity $V$ can be derived.
With the known $V$ we can define the correlation function of magnetic fluctuations $Q$ from Eq. (\ref{b_corr}).
Below we obtain this function for the stationary case when $dQ/dt=0$.

Here we use dimensionless units for the magnetic field strength:
\begin{equation}
\bar{H}=\sqrt{\frac{3}{\pi R v_0\rho_i\mu_{in}}}H\,.
\end{equation}
Here $v_0$ is the the velocity of turbulent motions in the scale $L\leq R$.

If dimensionless large scale magnetic field $\bar{H}_0<<1$, the energy of magnetic field fluctuations is concentrated
near the correlation length, $L_{corr}$, which is much smaller than the cloud size $R$.
Then the amplitude of magnetic fluctuations, $\delta H$, is much larger than $H_0$ \citep[see][]{kiselev13},
\begin{equation}
 \delta H\gg H_0\,,
\end{equation}
(cf. Equation~(\ref{H_res})).

By definition the correlation length is
\begin{equation}
L_{corr}\simeq\frac{1}{b_0^2}\int\limits_0^\infty \langle b_i(\textbf{0})b_i(\textbf{r}) \rangle dr\,,
\label{lcorr}
\end{equation}
and can be estimated if the function $Q$ is known from Equation~(\ref{b_corr}).

Then from numerical calculations of Equation~(\ref{lcorr}) we obtain that
\begin{eqnarray}
&&L_{corr} / R = 0.9 \bar{H}_0\,,
\label{L_ratio}
\\
&&\delta \bar{H} =  1.4 \sqrt{\bar{H}_0}\,.
\label{H_ratio}
\end{eqnarray}

We apply this theory to the molecular cloud Sgr B2. For estimates we use the following parameters of the cloud and the intercloud medium.
The gas density in the cloud $n_H\simeq 10^5$ cm$^{-3}$, the ionization degree in the dense molecular clouds $n_{H_3^+}/n_{H_2}\sim 10^{-8}$
\citep[see e.g.][]{oka06}, the radius of the Sgr B2 region emitting  gamma-rays is $R = 7$ pc
\citep[see][]{yang2}, the total magnetic field strength in Sgr B2, $H_0+\delta H \simeq 550$ $\mu$G \citep[][]{crutcher,crutcher99,crutcher10},
the average large scale magnetic field in the GC is about $H_0\sim 50-100$ $\mu$G \citep[][]{crock10}.
For the core Sgr B2, where the gas density is $n_H\sim 10^5$ cm$^{-3}$, and $\delta H/H_0\simeq 5.5$,
Equation~(\ref{L_ratio}) gives $L_{corr}\simeq 0.4$ pc, and Equation~(\ref{H_ratio}) gives $v_0 = 7.3$ km s$^{-1}$.
This estimates of the turbulent velocity is close to that presented in \citet{crutcher99} for Sgr B2.
In the outer envelope with $n_H\sim 10^3 - 10^4$ cm$^{-3}$, we expect that $\delta H\ll H_0$.

In a magnetic field $H=0.55$ mG the Larmor radius, $r_L$, of particles with energies $E<10^8$ GeV is smaller than $L_{corr}$.
In this medium propagation of magnetized particles along tangled magnetic field lines, can be described as diffusion with the
coefficient \citep[see for details][]{dog87}
\begin{equation}
D_c(E)\sim\frac{c\beta(E)L_{corr}}{2}\sim cL_{corr}\sqrt{\frac{(E/m_pc^2)^2+2(E/m_pc^2)}{(E/m_pc^2)^2+2(E/m_pc^2)+1}}\,,
\label{dc}
\end{equation}
where $\beta(E)=u/c$ and $u$ is the particle velocity . It follows from Equation~(\ref{dc}) that
\begin{equation}
D_c(E) \propto \left\{
\begin{array}{ll}
\sqrt{E}\,, & {\rm if}\quad  u<c \,,\\
{\rm constant}\,, & {\rm if}\quad  u\sim c \,,
\end{array}
\right.
\label{D_cloud}
\end{equation}
(cf. Equation~(\ref{D_st})).

\section{Spectrum of CRs inside and outside Sgr B2}\label{CR_Sgr B2}

The background hydrogen in the GC is ionised by subrelativistic CRs
\citep[see][for equations describing ionization processes see, e.g., \citet{dog13,dog14}]{oka05}.
The ionization rate of hydrogen, $\zeta$, is
\begin{equation}
\zeta =\int \sigma_HuN(E) dE \,,
\label{ionf}
\end{equation}
where $\sigma_H$ is the ionization cross-section of the molecular hydrogen by proton impact \citep[see][]{rudd,tati03},
and $u$ is the velocity of CR particles and $N(E)$ is their spectrum.

As we mentioned in Section~\ref{intro} the ionization rate in the diffuse molecular gas, $\zeta_0$ is about $3\times 10^{-15}$ s$^{-1}$,
while inside the core of Sgr B2 the rate, $\zeta_c$ is one order of magnitude smaller.
\citet{dog11,dog13,dog14} and \citet{tati12} presented arguments in favour of hydrogen ionization by subrelativistic electrons,
while the alternative process of ionization by high energy electrons was discussed in \citet{yusef13}. Below we discuss these models.

For calculations we approximate the gas distribution in Sgr B2 according to Equation~(\ref{nH2a}) or (\ref{nH2}):
\begin{itemize}
\item Model I: Sgr B2 consists of the two components as it was derived by \citet{lis91}, a dense core
with density $n_c\sim 10^5$ cm$^{-3}$ and radius $r_0=5$ pc surrounded by a diffuse component with
density $n_d\sim 10^3-10^4$ cm$^{-3}$ and radius $R=20$ pc;
\item Model II: The mass of the cloud is concentrated in a dense core  (as it follows from \citet{protheroe})
with density $n_c\sim 10^5$ cm$^{-3}$ and radius $r_0=5$ pc without an outer envelope.
\end{itemize}
For both distributions the total mass of the cloud is between $(2-6)\times 10^6$ M$_\odot$ and the gas column density in the
direction of the cloud center is $L_H\sim 10^{24}$ cm$^{-2}$ as follows from observations.

From Section~\ref{diffusion} it is clear that magnetic fluctuations with $\delta H> H_0$ are excited in the dense core
where the kinetic energy of turbulent motions is high enough. Only there we expect diffusion propagation of CRs with the coefficient (\ref{D_cloud}).
In the envelope, particles propagate without significant scattering because the fluctuations are damped by ion-neutral friction.

We assume that the spectrum of subrelativistic protons, $\bar{N}(E)$, in the intercloud medium can be presented as power-law,
\begin{equation}
\bar{N}(E)=K\left(\frac{E}{E_0}\right)^{\delta}\theta(E_{\rm max}-E)\,,
\label{n0}
\end{equation}
where $E$ is the kinetic energy of subrelativistic protons, $E_{\rm max}$ is the maximum energy of subrelativistic protons in the
intercloud medium and $E_0$ equals e.g. 1 MeV. Then from the ionization rate, $\zeta_0=3\times 10^{-15}$ s$^{-1}$, measured by
\citet{oka05} in the diffuse molecular clouds (outer envelope) and the rate of ionization inside Sgr B2,
$\zeta_c=3\times 10^{-16}$ s$^{-1}$ \citep[][]{vandertak}, we can estimate parameters of the CR spectrum (\ref{n0}).

The  magnetic fields are nonuniform inside molecular clouds and their strength correlates with the gas density as
\begin{equation}
H\propto n_H^{0.65}\,,
\end{equation}
see \citet{crutcher10}. We cannot say whether it is due to the frozen effect of the magnetic fields into the gas only or
this may be due to more effective generation of magnetic fluctuations in dense cores by the neutral gas turbulence as we assumed above.

\citet{padovani11} assumed a convergent structure of magnetic field line nearby molecular clouds and analysed the effect of CR mirroring.
They showed that this effect reduced the CR density in the clouds and the ionization rate there by a factor of $\sim 4$.

In the following we neglect variations of the large scale magnetic field in the outer envelope. We assume that inside the outer envelope
subrelativistic CRs propagate through the medium without significant scattering, i.e., along field lines of the large scale
magnetic field $H_0$ with their own velocity.
This is a big simplification of the process, but it provides an upper limit for the density of CRs penetrating into the envelope.
In this work we consider stationary distribution of CRs in molecular clouds.
Unlike the case of the source J1745.6$-$2858 near Sgr A* (right at the GC),
we do not have evidences indicating a time varying CR source at or near the location of Sgr B2.
Nevertheless, one can find a solution for the time-dependent case in \citet{chern14}.

We describe CR propagation inside the outer envelope as
\begin{equation}
c\beta(E)\frac{\partial N}{\partial x}-\frac{\partial }{\partial E}\left[\left(\frac{dE}{dt}\right)_i N\right]=0\,,
\label{equ}
\end{equation}
with the spectrum of CRs at the outer boundary of the envelope ($x=0$) as presented by (\ref{n0}).
The goal is to calculate the spectrum of subrelativistic CRs that reach the surface of the core at $r_0=5$ pc.

The solution of Eq.(\ref{equ}) is
\begin{equation}
N(x,E)=K\left(\frac{E}{E_0}\right)^{1/2}\left[\left(\frac{E}{E_0}\right)^{2}+\frac{ax\sqrt{m}}{E_0^2}\right]^{\frac{\delta+0.5}{2}}
\theta(E_{\rm max}^2-E^2-ax\sqrt{m})\,.
\end{equation}
Here we presented the term of energy losses in Eq.(\ref{iot}) as $dE/dt=a/\sqrt{E}$, and the mean free path of
a subrelativistic particle with mass $m$ and energy $E$ is $x\sim E/(dE/dt)_i\sqrt{2E/m}=E^2/a\sqrt{m}$.

The characteristic lifetime of a subrelativistic proton with energy $E$ is
\begin{equation}
\tau=\int\frac{dE}{(dE/dt)_i}=\frac{1}{3}\frac{m_e E^{3/2}}{\sqrt{2}\pi n_He^4\sqrt{m}\Lambda}\,.
\end{equation}

The energy loss gas column density, which a subrelativitic proton with energy $\bar{E}$ passes through during its lifetime, is
\begin{equation}
L_H=\frac{1}{3}\frac{m_e}{m}\frac{\bar{E}^{2}}{\pi e^4\Lambda}\,.
\end{equation}

\begin{figure}[ht]
\begin{center}
\includegraphics[width=0.7\textwidth]{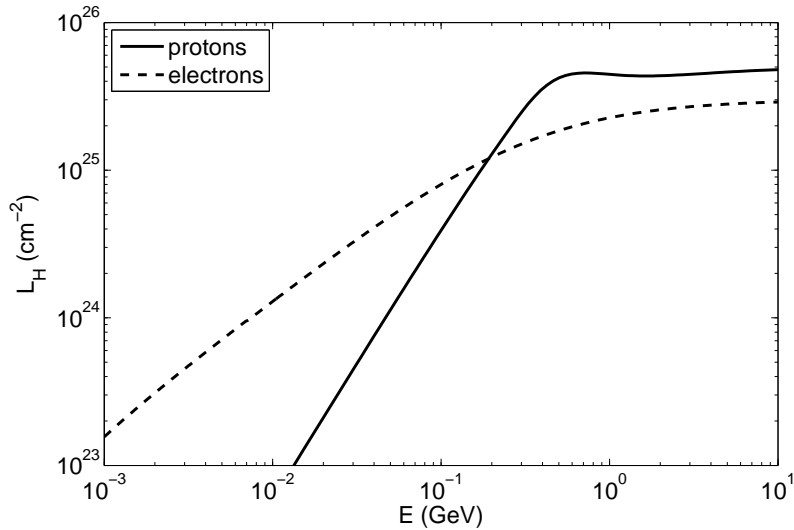}
\end{center}
\caption{Energy loss gas column density, $L_H$, passed by a proton (solid line) and an electron (dashed line) with energy ${E}$.}
\label{LH}
\end{figure}

In Figure~\ref{LH} we show the energy loss gas column density for electrons and protons derived from equations of energy losses taken from
\citet{haya,ginz} and \citet{mann94}.
For the fixed gas column density of the outer envelope taken from \citet{lis89,lis90,lis91},  which is presented by Equation~(\ref{nH2a}),
we can estimate the energy of protons which can reach the surface of the central core.
For $L_H\simeq 10^{23}$ cm$^{-2}$ this value is about $E\ga 20$ MeV .

\begin{figure}[ht]
\begin{center}
\includegraphics[width=0.7\textwidth]{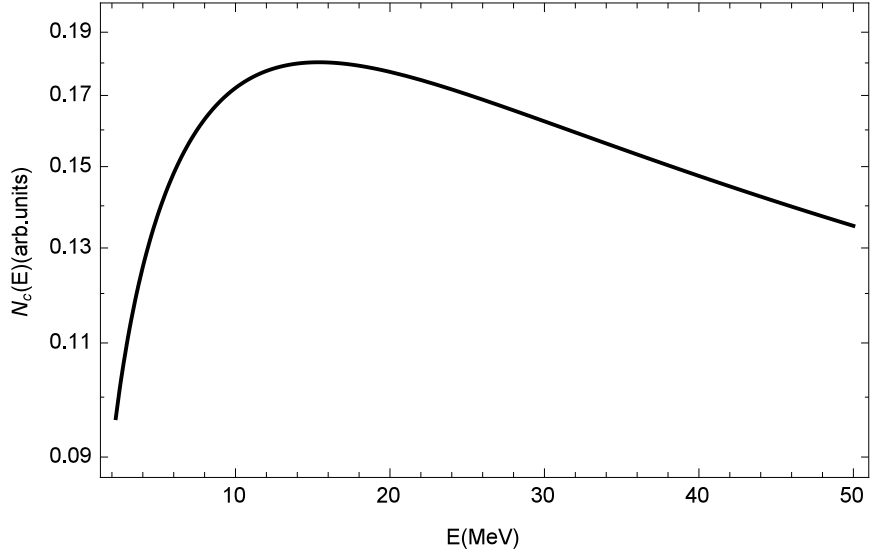}
\end{center}
\caption{Expected spectrum of proton at the boundary of the central core $N_c(E)$ for  $L_H= 10^{23}$ cm$^{-2}$
and the spectrum of protons in the intercloud medium $\bar{N}(E)=K(E/E_0)^{-0.5}$.}
\label{spectrum1}
\end{figure}

As an example, in Figure~\ref{spectrum1} we show the expected spectrum of protons at the boundary of the central core $N_c(E)$
for $L_H= 10^{23}$ cm$^{-2}$ if the spectrum of protons in the intercloud medium is $\bar{N}(E)=K(E/E_0)^{-0.5}$.

According to Section~\ref{diffusion}, inside the core CR propagation is described as diffusion with the coefficient \citep[see][]{dog87}
\begin{equation}
D_c=\frac{uL_{corr}}{2}\,,
\end{equation}
where $L_{corr}=0.4$ pc.

The distribution function of protons, $N_c(E)$, is described by the equation
\begin{equation}
-\frac{1}{r^2}\frac{\partial }{\partial r}\left(r^2D_c\frac{\partial {N}_c}{\partial r}\right)
-\frac{\partial }{\partial E}\left[\left(\frac{dE}{dt}\right)_i{N}_c\right]=0\,,
\label{equ1}
\end{equation}
where $(dE/dt)_i$ is the rate of ionization losses in the dense core, calculated for the density $n_H\sim 10^5$ cm$^{-3}$.

The boundary condition on the core surface, $r=r_0$
\begin{equation}
N_c(E)=K\left(\frac{E}{E_0}\right)^{1/2}\left[\left(\frac{E}{E_0}\right)^{2}+\frac{a(R-r_0)\sqrt{m}}{E_0^2}\right]^{-\frac{\delta+0.5}{2}}
\theta(E_{\rm max}^2-E^2-a(R-r_0)\sqrt{m})\,,
\end{equation}
where $R$ and $r_0$ are the radius of the outer envelope and the central core (see Model I).
We can ignore the unknown constant $K$ from the ratio of ionization rates outside and inside the core.

The ionization rate averaged over the scale $L_H$ of the outer envelope is $\zeta =3\times 10^{-15}$ s$^{-1}$,
\begin{equation}
\zeta_c \simeq\frac{1}{L_H}\int\limits_R dr \int \sigma_HuN_c(E,r) dE \,.
\label{ionf1}
\end{equation}
The ionization rate inside the central part of core is $\zeta_c=3\times 10^{-16}$ s$^{-1}$. Then
\begin{equation}
\frac{\zeta}{\zeta_c}\simeq\frac{r_0}{(R-r_0)}\frac{\int\limits_{r_0}^{R} dx
\int\limits_{E_{\rm min}}^{E_{\rm max}}dEN(x,E)\sigma_i(E)u(E)}{\int\limits_0^{r_0} dr
\int\limits_{E_{\rm min}}^{E_{\rm max}}dEN_c(r,E)\sigma_i(E)u(E)}\,.
\label{ratio}
\end{equation}
From Equation~(\ref{ratio}) we can derive the spectral index $\delta$ as a function of $E_{\rm max}$ for a given value of $L_H$ in the outer envelope,
where $L_H=n_H(R-r_0)$, $R=20$ pc and $r_0=5$ pc. The constant $K$ is calculated then from the ionization rates inside or outside the cloud.

\begin{figure}[ht]
\begin{center}
\includegraphics[width=0.7\textwidth]{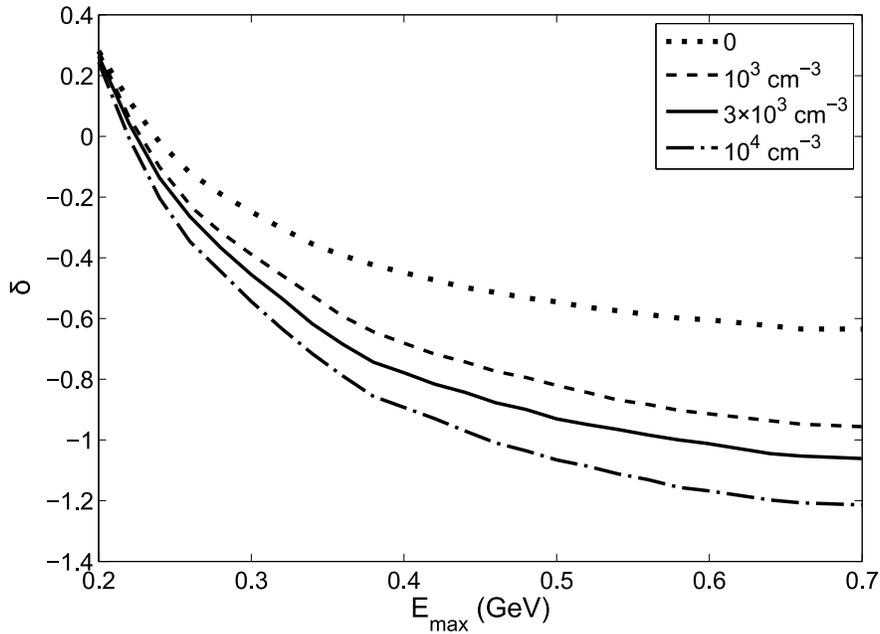}
\end{center}
\caption{The required maximum energy of protons $E_{\rm max}$ as a
function of the spectral index $\delta$ and the  gas density in
the outer envelope, when $n_H\neq 0$ in the outer envelope (Model I), and for $n_H=0$ in the outer envelope (Model II).}
\label{emax}
\end{figure}

In Figure~\ref{emax} we show the spectral index $\delta$ for different values of $n_H$.
The case $n_H=0$ corresponds to the case where the whole mass of Sgr B2 is concentrated in the core (i.e. Model II).
One can see from the figure that for very hard spectra with $\delta=0.2$ almost all particles reach  the core surface.
The required maximum energy $E_{\rm max}$ of protons is in this case about $0.2$ GeV independent of the gas density.

\begin{figure}[ht]
\begin{center}
\includegraphics[width=0.7\textwidth]{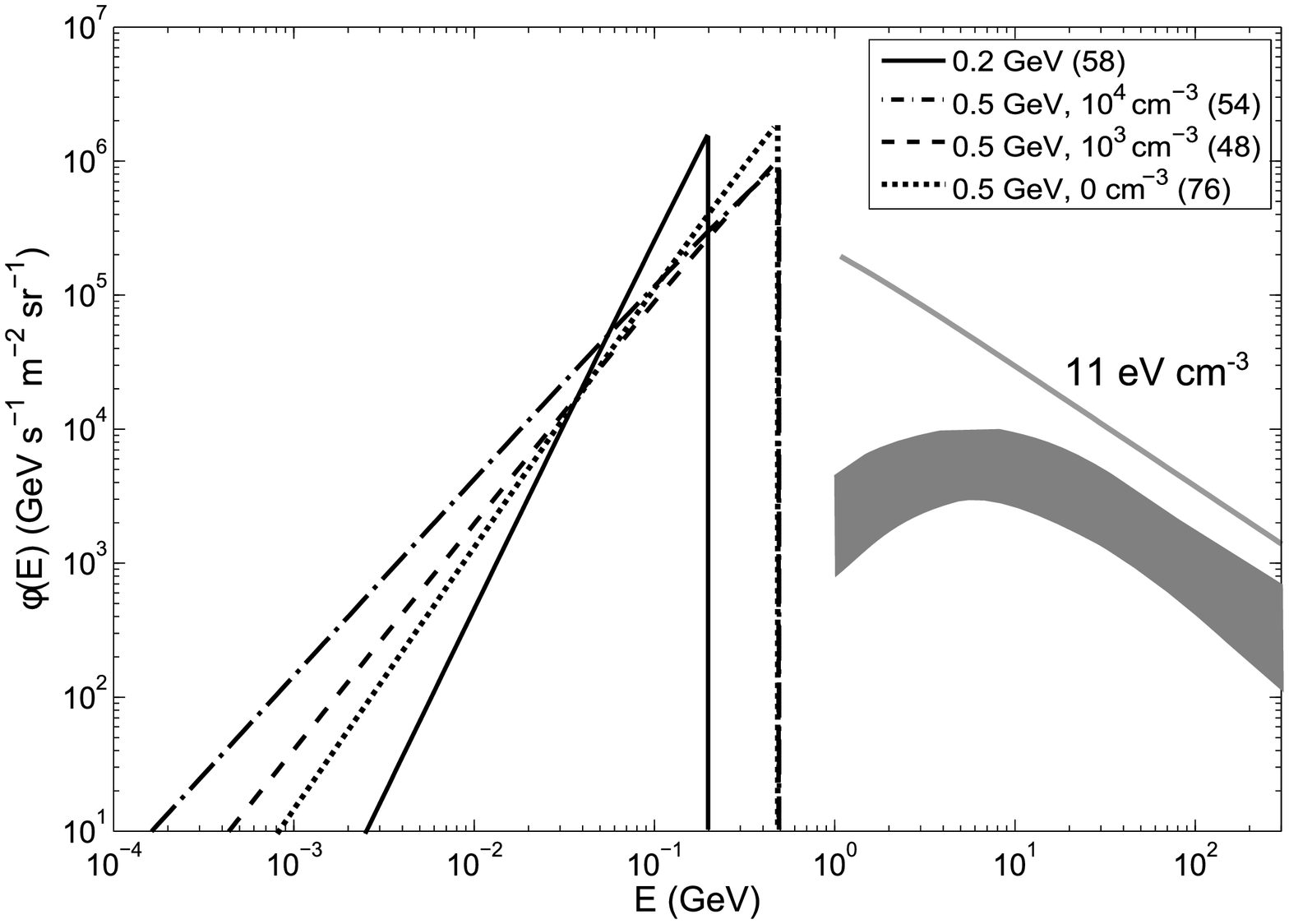}
\end{center}
\caption{Spectrum of subrelativistic protons in the intercloud medium, needed for the rate of ionization inside and outside
Sgr B2 for different values of $E_{\rm max}$. The number in the brackets represent the estimated energy density of subrelativistic
protons in eV cm$^{-3}$. The solid line (left side of the figure) represents spectra of subrelativistic protons for  hard spectra  with
$\delta=0.2$ that gives the value of $E_{\rm max}=0.2$ GeV (see Fig. \ref{spectrum1}).
In this case  almost all particles freely penetrate through the envelope, and therefore the the required spectrum of subrelativistic protons
is the same independently of $n_H$. The other three curve are shown for $E_{\rm max}=0.5$ GeV and different values of $n_H$.
As follow from Figure~\ref{spectrum1} for this value of $E_{\rm max}$ the spectra are relatively soft and therefore a part of particle
is absorbed in the outer envelope. This makes the result depended on $n_H$ as shown in the figure.
The spectra of relativistic protons are shown by shaded area for the case of outer envelope and the line
when the mass is concentrated in the core. Comments are presented in the text.}
\label{psp}
\end{figure}

The derived spectrum of subrelativistic protons is shown in Figure~\ref{psp} for different values of $n_H$ in the envelope and $R-r_0=15$ pc.
The derived values of $E_{\rm max}$, the density of hydrogen $n_H$, and the energy density of subrelativistic  protons needed for the
ionization rate in Sgr B2 are also shown in the figure.

The relativistic component of CR spectrum inside Sgr B2 can be derived from the observed flux of gamma-rays there \citep[see][]{yang2}.
Gamma-ray flux generated by relativistic protons in molecular clouds can be calculated by
\begin{equation}
F_\gamma(E_\gamma) = 4\pi c\int\limits_0^Rn_H(r)r^2dr\int_E N(E,r)
\frac{d\sigma}{dE_\gamma}(E,E_\gamma)dE\,,
\label{fgamma}
\end{equation}
where $n_H(r)$ is the hydrogen distribution, $d\sigma/dE_\gamma(E,E_\gamma)$ is the differential cross-section for gamma-ray production in
proton-proton collisions \citep{kamae}, and  $N(E,r)$ is the distribution of relativistic protons in Sgr B2.

If the mass of Sgr B2 is concentrated in the envelope (i.e. Model I), then according to our analysis relativistic
protons fill uniformly this region and their density is the same as in the intercloud medium.
It is clear from Figure~\ref{spectrum1} that even protons with $E\simeq 0.2$ GeV can penetrate through the outer envelope.
Protons with energies $E\ga 1$ GeV pass through the envelope without absorption ($\tau_i,\tau_{pp}>\tau_{pr}$).
The spectrum derived for this case by \citet{yang1,yang2} is shown as shaded areas in Figure~\ref{psp}.
Variations of the spectrum are governed by the uncertainties of the mass of Sgr B2, which is in the range of $2-10\times 10^6$ M$_\odot$.
The energy density of relativistic protons is then within 1 to 4 eV cm$^{-3}$.

It is interesting to notice that in this case the required energy density of subrelativistic protons is more than one order
of magnitude higher than that of relativistic protons.

If the mass of Sgr B2 is mainly concentrated in the core (i.e. Model II), the distribution of relativistic protons in the cloud is calculated
similarly to Equation~(\ref{equ1}) ,
\begin{equation}\label{rpr_cl}
\frac{N_c}{\tau_{pp}(E)} -  \frac{1}{r^2}\frac{\partial}{\partial r}\left(r^2D_c\frac{\partial}{\partial r} N_c\right)= 0\,,
\end{equation}
where $D_c\sim cL_{corr}/2=$constant and $\tau_{pp}=(n_Hc\sigma_{pp}(E))^{-1}$ is the characteristic time of $pp$ collisions.
The boundary condition is
\begin{equation}
N_c(E)\left|_{r=r_0}\right.=\bar{N}(E)\,,
\end{equation}
where $\bar{N}(E)$ is the spectrum of relativistic protons in the intercloud medium, which is supposed to be a power-law.

In this case protons are distributed in the core nonuniformly ($\tau_{pp}<\tau_{pr}$) and their spectrum can be derived from
Equations~(\ref{fgamma}) \& (\ref{rpr_cl}). Results of the calculations are shown as the solid line in Figure~\ref{psp}.
One can see that the energy density of relativistic protons is about one order of magnitude higher than that near Earth,
but it is still smaller than that of subrelativistic CRs.

From Figure~\ref{psp}, it is reasonable to conclude that relativistic and subrelativistic protons have different sources in the GC
because their spectra have a gap in intensities in the range about several hundred MeV
and they do not match smoothly with each other at these energies.
An exception is the case where the whole mass of Sgr B2 is concentrated in the core
(the solid line in the subrelativistic energy range and the straight line in the relativistic energy range).
It seems from the figure that in this case the distribution of subrelativistic and relativistic protons can be described by a unified
spectrum with a break of the spectral index from $\delta\simeq -3$ in the relativistic energy
to $\delta\simeq 0.2$ in the subrelativistic energies.
Similar spectrum can be formed in the intercloud space, e.g., by ionization losses (see Equation~(\ref{iot}))
if the spectrum of protons injected by sources has a cut-off, $Q(E)=KE^\delta \theta(E-E_{\rm min})$.
For the equation
\begin{equation}
\frac{d}{dE}\left[\left(\frac{dE}{dt}\right)_iN(E)\right]=Q(E)\,,
\end{equation}
the solution for $E<E_{\rm min}$ is
\begin{equation}
N(E) \propto E^{0.5} \,.
\label{ion_sp}
\end{equation}
For more details of this solution see \citet{dog09b}.

\section{ By-products of the model of hydrogen ionization by subrelativistic protons}

With the spectrum of protons derived in the previous section we can analyse whether this model can be proved or disproved from observations.

\subsection{Emission of the 6.4 keV iron line}

First we would like to mention that subrelativistic protons not only ionize hydrogen molecules in the clouds but
also atoms of heavier elements, e.g., iron, whose ionization of K-$\alpha$ shell is observed as the 6.4 keV line emission.
It was proved that this emission is generated by a front of hard X-ray photons emitted in the past by Sgr A* \citep[see][]{koyama96}
whose luminosity was much higher in the past than at present \citep[see e.g.][]{ryu13}.
The emission shows time variability \citep[see, e.g.,][]{koyama08}, and it had decreased by one order of magnitude from the peak period
in 2000 to 2013 \citep[][]{inui09, nobu11,nobu14}.
It is expected that for about twenty years of observations of Sgr B2, the front of X-rays have to leave (almost) the cloud.
Then we may expect that a background flux of the line emission from Sgr B2 generated by  CRs can be seen at present or in the near future
\citep[see][]{dog11}.

Temporal variations of the 6.4 keV Sgr B2 flux in the period 2005-2013 are presented in Table \ref{tbl-2}.
\begin{table}[h]
%\begin{center}
\caption{The 6.4 keV flux from Sgr B2 region of  $2.0^\prime$ radius.}
\label{tbl-2}
\begin{tabular}{cccccccccccccccccccc}
\tableline\tableline
&&&&Year &&&&&&&&&&& FeI-K$\alpha$\tablenotemark{a,b}&&&&\\
\tableline
&&&&2005&&&&&&&&&&&$(18.0\pm 1.3)\times 10^{-5}$&&&& \\
&&&&2009 &&&&&&&&&&&$(7.4\pm 0.8)\times 10^{-5}$&&&& \\
&&&&2013 &&&&&&&&&&&$(3.0\pm 0.9)\times 10^{-5}$  &&&& \\
%&&&&2013 &&&3.0&&& $4.5\pm 0.47$ &&& \\
\tableline
\end{tabular}
%% Any table notes must follow the \end{tabular} command.
\tablenotetext{a}{Errors at 90\% confidence levels. Background is taken into account.}
\tablenotetext{b}{Observed flux in the unit of photons cm$^{-2}$ s$^{-1}$.}
\tablecomments{The data were derived from \citet{nobu11,nobu14}.}
%\end{center}
\end{table}

Below we compare these Suzaku data with model estimations.
From the spectrum shown in Figure~\ref{psp} we can estimate the expected 6.4 keV line emission generated by the protons:
\begin{equation}
I_{6.4} = \frac{1}{R_{GC}^2} \int\limits_0^R n_H(r)r^2dr \int\limits_E N(E,r) v\sigma_{6.4}dE\,,
\end{equation}
where the cross-section of 6.4 keV photon  production by proton impact $\sigma_{6.4}$ is from \citet{tati03}.

The expected intensity depends weakly on model parameters and is about
\begin{equation}
I_{6.4} \approx (3-5)\times 10^{-6} \eta\ {\rm ph}\ {\rm cm}^{-2}\ {\rm s}^{-1}\,,
\label{6.4}
\end{equation}
where $\eta$ is the abundance of iron atoms with respect to solar.
The estimated iron abundance $\eta$ in the Sgr B2 is in the range $1.3 - 1.9$
\citep[see, e.g.,][]{revn04,nobu11}. The estimated flux of the 6.4 keV line in Eq. (\ref{6.4}) is three times smaller than it was in 2013.
Hence in the framework of the model, subrelativistic protons might produce in 2013 about 30\% of the total flux from Sgr B2.
We cannot exclude that at present, the front of hard X-rays has left Sgr B2 and this emission is generated by subrelativistic CRs.
Then it should drop down to the level predicted by Eq. (\ref{6.4}).
Future analysis of the X-ray data for the period after 2013 may confirm our conclusion.
Restrictions on processes of 6.4 keV line production can be obtained from measurements of the line broadening \citep[see][]{dog98}.
We hope that this crucial information to constrain the origin of the 6.4 keV emission from Sgr B2 will be obtained by
ASTRO-H \citep[see][]{koyama14}.
% The expected equivalent width of the 6.4 keV iron K$\alpha$ line produced %solely by CR protons is $0.7\eta$ keV.

\subsection{Hard X-ray continuum}

Hard X-ray photons and subrelativistic CRs also produce a flux of continuum emission by Thomson scattering of  photons (Compton echo)
or by bremsstrahlung losses of charged particles.
Observations presented by \citet{terrier10} showed time variability of the X-ray flux from Sgr B2 in the range from 20 to 100 keV which
correlated nicely with the variability of the 6.4 keV line emission from Sgr B2.
This confirmed that their common origin is due to X-ray irradiation from Sgr A*.

The X-ray emission produced by inverse bremsstrahlung of protons can be calculated from
\begin{equation}
I_{x}(E_x) = \frac{1}{R_{GC}^2} \int\limits_0^R n_H(r)r^2dr \int\limits_E N(E,r) \frac{d\sigma_x}{dE_x}(E,E_x) dE\,,
\label{br_i}
\end{equation}
where $\sigma_x/dE_x$ is the cross-section of inverse bremsstrahlung
\begin{equation}
\frac{d\sigma_x}{dE_x}(E,E_x)=\frac{8}{3}\frac{e^2}{\hbar c}\left(\frac{e^2}{m_ec^2}\right)^2
\frac{m_e c^2}{\hat{E}E_x}\ln\left[\frac{\left(\sqrt{\hat{E}}-\sqrt{\hat{E}-E_x}\right)^2}{E_x}\right]\,,
\label{br_cross}
\end{equation}
where $\hat{E}=(m_e/m_p)E$ for protons and  $\hat{E}=E$ for electrons.
Below we adopt the cross-section from \citep{haug97}.
Subrelativistic protons with the energy $E$ emit bremsstrahlung photons with energies $E_x\la (m_e/m_p)E$ \citep[see][]{haya}.
Therefore for the cutoff energy of subrelativistic protons  about $0.2-0.5$ GeV
we expect that they generate hard X-ray emission in the range $E_x\la 10-30$ keV.

The calculated X-ray fluxes depending of the Sgr B2 mass are about
\begin{eqnarray}
&I_x^{2-10} =(4-6)\times 10^{-13}\ {\rm erg}\ {\rm s}^{-1}\ {\rm cm}^{-2} &
{\rm for}\quad 2\ {\rm keV} \leq E_x \leq 10\ {\rm keV}\,,
\label{2-10}
\\
&I_x^{20-60} = 7.5\times 10^{-13}\ {\rm erg}\ {\rm s}^{-1}\ {\rm cm}^{-2} &
{\rm for}\quad 20\ {\rm keV} \leq E_x \leq 60\ {\rm keV}\,.
\label{20-100}
\end{eqnarray}
One can see that the model estimations are significantly below the experimental data presented in \citet{nobu11} for the range
$2-10$ keV and in \citet{terrier10} for the range $20-60$ keV.
However, the continuum X-ray flux shows strong decrease in the Suzaku (2-10 keV) and the INTEGRAL (20-100 keV) ranges.
For the period from 2003 to 2009 the flux dropped down by a factor 0.4 \citep[][]{nobu11, terrier10}.
Thus, one can expect that the continuum emission produced by subrelativistic CRs (Equations~(\ref{2-10}) \& (\ref{20-100}))
may be seen in near future.

\subsection{De-excitation nuclear lines}

Subreletivistic protons and nuclei with energies between a few MeV and several hundred MeV generates in the interstellar medium
emission of nuclear gamma-ray lines in the range from 0.1 to 10 MeV \citep[see the review of][]{tati03}.
Estimations of \citet{dog09b} and \citet{ben13} showed, however, that this line emission cannot be detected even in the GC
because the sensitivity of existing gamma-ray telescopes is not high enough.
It seems nevertheless that the discovery of the gamma-ray lines will be possible with new generation telescopes.
Bearing this in mind, we estimated the flux of C and O lines from Sgr B2, expected for the spectrum of subrelativistic protons,
derived in Section~\ref{CR_Sgr B2}.
For the double solar abundance in the GC the flux of the O line from there is about $1.38\times 10^{-7}$ ph cm$^{-2}$ s$^{-1}$
and the C line $8.34\times 10^{-8}$ ph cm$^{-2}$s$^{-1}$.

\subsection{Secondary electrons and radio emission from Sgr B2}

Relativistic protons generate from $pp$ collision not only gamma-rays, as discussed in Section~\ref{CR_Sgr B2}, but also
secondary electrons, which should produce a synchrotron radio emission in the relatively strong magnetic fields of Sgr B2.
From the observed gamma-ray flux we can estimate the density of secondary electrons and the flux of radio emission using the
equations for these processes presented, e.g., in \citet{ber90}.
The flux of radio emission from Sgr B2 in the range 330 MHz to 2368 MHz was presented in \citet{jones} who concluded that they did not
find any evidence for non-thermal emission from Sgr B2 and the observed radio flux from there is of the thermal origin.

\begin{figure}[ht]
\begin{center}
\includegraphics[width=0.6\textwidth]{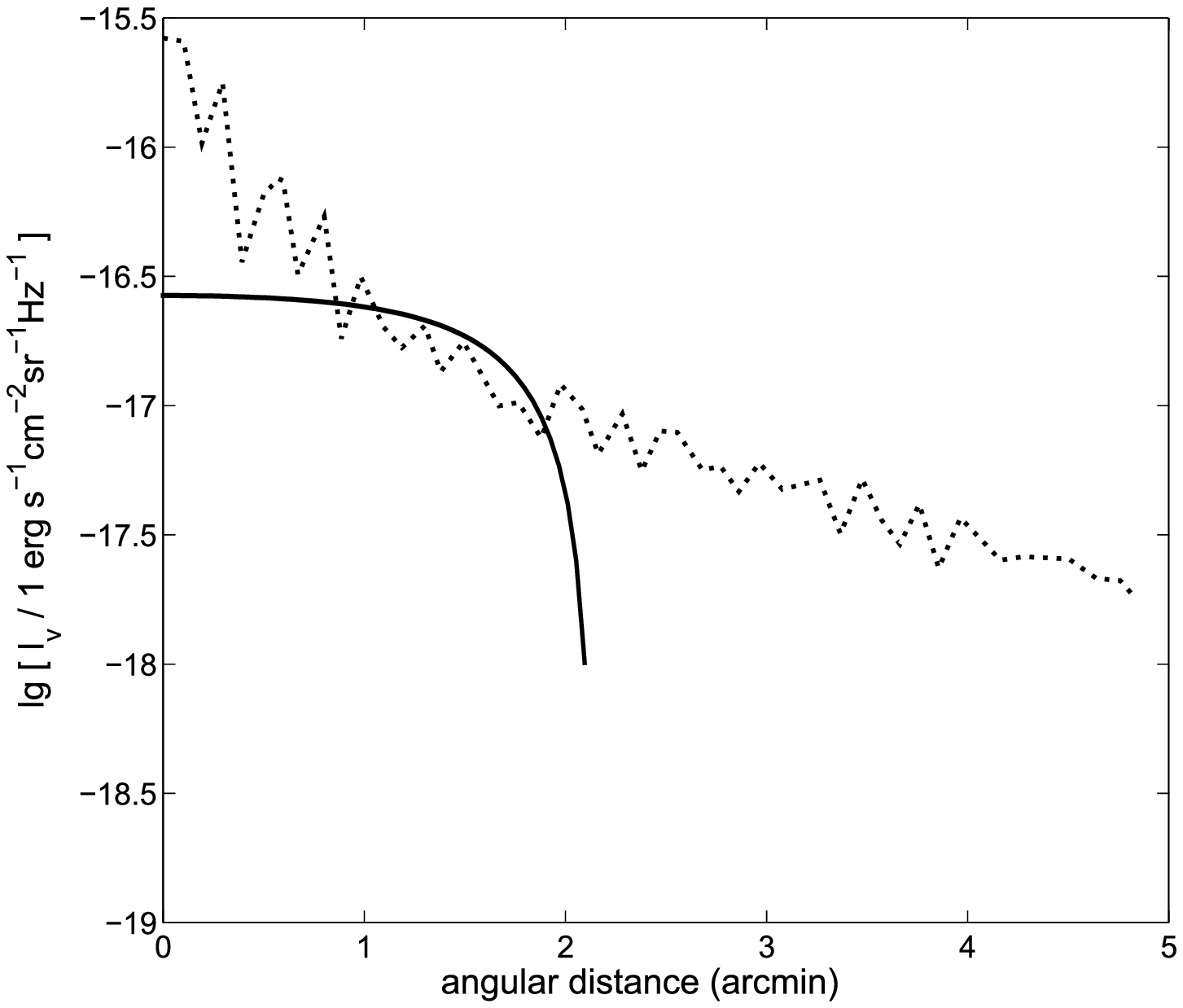}
\end{center}
\caption{Upper limit on the angular distribution of the radio emission at 330 MHz (solid line)
together with experimental data from \citet{jones} (dotted line)}\label{spectrum}
\end{figure}

We estimated the synchrotron component generated by secondary electrons.
As an upper limit for the density of secondary electrons we derive their spectrum from the gamma-ray data from Sgr B2
whose mass is $2\times 10^6$ M$_\odot$.
The derived angular distribution of the 330 MHz emission for this case is shown in Figure~\ref{spectrum}
together with experimental data taken from \citet{jones}.

%References for radio from Sgr B2 \citep{protheroe,jones,dickin}
We see that even for the most favourable parameters of the model the nonthermal  component is below the intensity of thermal emission.

The flux of nonthermal component can be subtracted from the total radio emission  from Sgr B2 by observation of polarization.
A flux of synchrotron emission is polarized if the strength of large scale magnetic fields is high enough \citep[see, e.g.,][]{ginz}.

However as we concluded in Section \ref{CRdiff} the chaotic component of the magnetic field, $\delta H$, is five times
larger than the large scale magnetic field, $H_0$, whose strength is estimated as 100 $\mu$G. The expected linear polarization
degree, $p_l$, was derived by \citet{kors}, which is
\begin{equation}
p_l=\frac{(\gamma+1)}{(\gamma+7/3)}\frac{(\gamma+3)(\gamma+5)}{32}
\left[1-\frac{(\gamma^2+8\gamma+3)}{24}\frac{H_0^2} {\delta H^2}\right]\frac{H_0^2}{\delta H^2}
\quad{\rm for}\quad\delta H\gg H_0\,.
\end{equation}

For the expected spectral index of secondary component $\gamma$ (for the electron spectrum $\propto E^{-\gamma}$) with $\gamma\sim 3-4$,
the estimated polarization degree of the nonthermal component of Sgr B2 is about $p_l\simeq 3.5-4.8$\%.

If, however, the turbulence inside the clouds is not strong enough and the amplitude of magnetic fluctuations is small ($\delta H\ll H_0$),
then the polarization degree is described by the equation \citep[see][]{kors}
\begin{equation}
p_l=\frac{(\gamma+1)}{(\gamma+7/3)}\left(1-\frac{2}{3}\frac{\delta H^2}{H_0^2}\right)
\quad{\rm for}\quad\delta H\ll H_0\,.
\end{equation}
The upper limit of linear polarization of synchrotron radiation can reach the level about $75-80$\%, if the scale of $H_0$ is about the cloud radius.

We hope that the next generation radio telescope SKA \citep[see][]{strong14,dickin} may be able to detect the nonthermal flux of Sgr B2.

\section{Hydrogen ionization by primary electrons in Sgr B2}

Alternatively ionization of the molecular gas may be caused by high energy electrons \citep[see][]{yusef13}.
Below we analyse whether the molecular hydrogen in Sgr B2 is ionized by electrons and whether we can distinguish
from observations between processes of proton and electron ionization there.

As follows from Figure~\ref{LH} subrelativistic CRs are absorbed in the outer envelope, and electrons with energies
above 1 MeV are able to penetrate into the dense core.
Therefore, for estimates of processes of hydrogen ionization and bremsstrahlung emission from there we can use equations
in the relativistic energy range, which simplifies the calculations significantly.
The intensity of bremsstrahlung emission is determined by a cutoff position derived from the rate of ionization losses in the outer envelope
(see, e.g., Figure~\ref{spectrum1}).

Electrons lose about 42 eV per one act of ionization \citep[see][]{dalg}. We we can estimate the ionization rate of hydrogen as
\begin{equation}
\zeta =\frac{n_e}{42\ {\rm eV}}\frac{dE}{dt}\,,
\end{equation}
where $n_e$ is the density of relativistic electrons in Sgr B2,  $dE/dt$ represents ionization losses of
these electrons per H-atom and per one electron.
We notice that in the relativistic energy range the rate of ionization losses weakly depends on the electron energy.
From the average ionization rate in Sgr B2, $\zeta_0 = 4\times 10^{-16}$ s$^{-1}$, we obtain that the required density
of electrons with any energy above 1 MeV is $n_e\simeq 7.7\times 10^{-8}$ cm$^{-3}$.

These electrons provide X-ray bremsstrahlung radiation. However, unlike the case of subrelativistic protons, the energy of
bremsstrahlung photons is $E_x\la E_e$.
The X-ray emissivity of relativistic  electrons with $E_e>1$ MeV can be estimated from \citet{haug98}.
For the range of 20-60 keV the bremsstrahlung emissivity $\epsilon_{ep} = 5\times 10^{-23}$ erg s$^{-1} $H$^{-1}$ per electron for scattering
by background protons and  $\epsilon_{ee} = 8.5\times 10^{-23}$ erg s$^{-1}$H$^{-1}$ for scattering on background electrons.
The total 20-60 keV X-ray flux from Sgr B2 is
\begin{equation}
F_x=\frac{n_e(\epsilon_{ep}+\epsilon_{ee})M_H}{4\pi m_pR_{GC}^2}
\approx 1.7\times 10^{-12}\ {\rm erg}\ {\rm s}^{-1}\ {\rm cm}^{-2}~\times\left(\frac{M_H}{10^6 {\rm M}_\odot}\right) \,,
\label{ee_xrays}
\end{equation}
where $M_H$ is the total mass of Sgr B2.

\begin{figure}[ht]
\begin{center}
\includegraphics[width=0.6\textwidth]{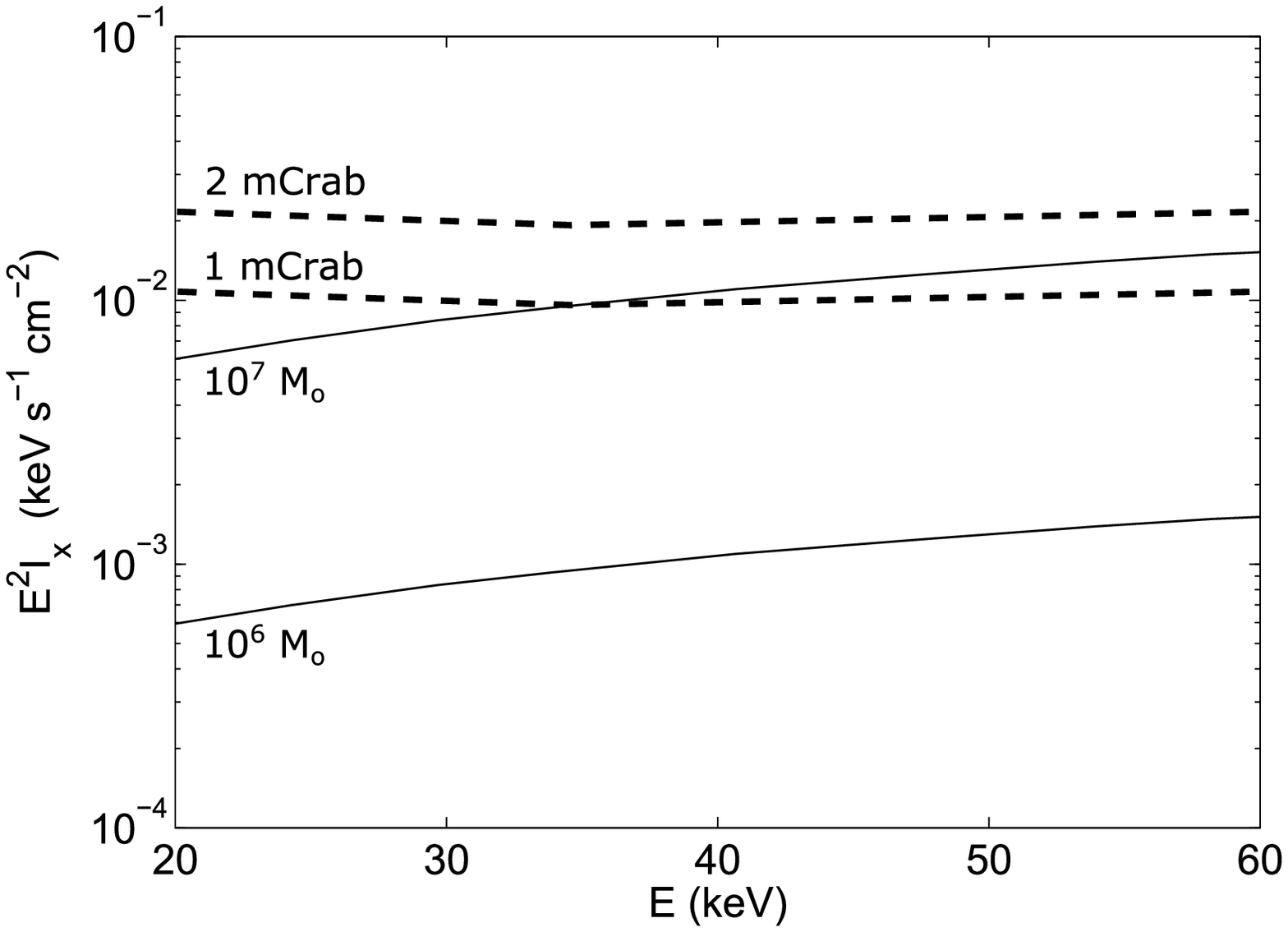}
\end{center}
\caption{Spectrum of hard X-ray emission observed by INTEGRAL from \citet{terrier10}.
The bremsstrahlung emission of 10 MeV electrons for upper $10^7$ M$_\odot$ and lower $10^6$ M$_\odot$
limits of the mass of Sgr B2 (solid lines). The X-ray spectrum form \citet{terrier10} is shown by dashed lines.}
\label{terrier}
\end{figure}

For the mass of Sgr B2 $M = 8\times 10^6$ M$_\odot$ this X-ray flux reaches the level of 1 mCrab.
Just this flux was observed by \citep{terrier10}.
As we noticed above the origin of time variable continuum X-ray emission is the irradiation of Sgr B2 by hard
X-ray emitted in the past by Sgr A*.
As follows from the observations of Suzaku and INTEGRAL \citep{nobu11,terrier10} the X-ray flux is a power-law with
$F_x\propto E_x^{-2}$.
The electron bremsstrahlung of relativistic electrons is hard, $\propto E_x^{-1}$, as it follows from Equations~(\ref{br_i}) \& (\ref{br_cross}).
Then the effect of electron bremsstrahlung is evident in the total X-ray flux from Sgr B2 at relatively high values
of $E_x$ as a stationary excess above $E_x^{-2}$, see Figure~\ref{terrier}.
We can conclude from our calculations that the origin of ionization in Sgr B2 can be, in principle, determined from the INTEGRAL observations.
If these observations show an excess of hard X-ray emission above the power-law $\propto E_x^{-2}$ at large energies,
then it is in favour of the electron origin of hydrogen ionization. If not, then only the proton origin is acceptable.

\section{Conclusion}
The conclusion of this paper can be itemized as follows:
\begin{itemize}
\item The total mass and the  gas distribution in the cloud Sgr B2 is highly uncertain.
Therefore we study two extreme cases: (1) when the mass is mainly distributed in the relative low density outer
envelope ($n_H\sim 10^3-10^4$ cm$^{-3}$) with the radius about 22 pc (Model I),
(2) when the most of the mass is concentrated in the dense core ($n_H\sim 10^5$ cm$^{-3}$) with the radius about 5 pc (Model II).
\item We conclude that propagation of high energy particle in dense molecular clouds is determined by the amplitude of
magnetic fluctuations generated by turbulent motions of neutral gas and their damping due to ion-neutral friction.
In the outer envelope of Sgr B2 (if it is) with the density about $10^3-10^4$ cm$^{-3}$ the kinetic energy of turbulent motions
is quite low and it does not generate fluctuations with amplitude higher than the strength of large scale magnetic field, $H_0\sim 100$ $\mu$G.
Therefore we assume that particles propagate their without scattering.
\item In the dense core with the density about $10^5$ cm$^{-3}$ the neutral gas turbulence generates fluctuations with the amplitude
$\delta H\gg H_0$. Our estimations show that $\delta H\sim 500$ $\mu$G for the parameters of Sgr B2 core.
Then, unlike previous models of CR propagation in molecular clouds, we conclude that this process is determined by chaotic magnetic fields
generated by turbulent motions of the molecular gas in the core.
The key parameter of this process is the correlation length of the chaotic field, $L_{corr}$ which determines diffusion propagation
of particles inside the core. For the magnetized particles the diffusion coefficient is constant in the relativistic energy range,
and is proportional to $\sqrt{E}$ in the subrelativistic energy range.
For the parameters of Sgr B2 the correlation length is about $L_{corr}\sim 0.4$ pc.
\item For the model of hydrogen ionization by subrelativistic protons we derived their spectrum depending on the hydrogen density
in the outer envelope. We showed that the hydrogen ionization is provided by protons with the energy above 20 MeV.
Their energy density is about or above 50 eV cm$^{-3}$.
\item The spectrum of relativistic protons was derived from the observed gamma-ray flux of Sgr B2.
If the mass of the cloud is mainly distributed in the outer envelope with the radius of 22 pc,
then the energy density of these protons is about the same as near Earth, i.e., $1-4$ eV cm$^{-3}$.
If, however, most of mass is concentrated in the core with the radius 5 pc then the required energy density of the protons is about 10 eV cm$^{-3}$.
\item We conclude that if subrelativistic protons are responsible for hydrogen ionization in Sgr B2,
they generate also the 6.4 keV line emission and continuum emission in the X-ray energy range $\la 30$ keV.
The expected flux in the K-$\alpha$ iron line from Sgr B2 is about $10^{-5}$ ph cm$^{-2} $s$^{-1}$,
which is only three times below the line flux observed in 2013 by Suzaku.
It was shown that the observed flux of the line from Sgr B2 was generated by X-ray photons emitted by Sgr A* in the past.
However, this flux is decreasing rapidly and we expect that in near future we might observe a component of the 6.4 keV emission
from Sgr B2 generated by CRs.
\item The expected continuum X-ray emission generated by proton bremsstrahlung is about $I_x^{2-10} =(4-6)\times 10^{-13}$ erg cm$^{-2}$ s$^{-1}$
in the range $2-10$ keV, and $I_x^{20-60} = 7.5\times 10^{-13}$ erg cm$^{-2}$ s$^{-1}$ in the range $20-60$ keV.
Both fluxes are below the continuum fluxes from Sgr B2 measured by Suzaku and INTEGRAL in 2009.
However we expect that the continuum X-ray fluxes might decrease to the level predicted by the model of proton bremsstrahlung.
It is highly desirable to get updated results from Suzaku and INTEGRAL.
\item Subrelativistic protons generate also de-excitation gamma-ray lines.
However, the estimated flux of the C and O lines from Sgr B2 is too low for detection by present gamma-ray telescopes.
We hope that it may be a target for future gamma-ray missions.
\item Relativistic protons in Sgr B2 produce secondary electrons by $pp$ collisions whose density can be derived from the observed gamma ray flux.
Thus, a flux on nonthermal synchrotron radio emission is expected from Sgr B2.
However, observations showed that the flux of thermal radio emission was much higher than the estimated flux of synchrotron emission.
Therefore, it is problematic to subtract the nonthermal component from the total Sgr B2 radio flux.
However, unlike the thermal component, the flux of synchrotron emission is polarized. We estimated the expected degree of polarization
of the nonthermal component.
If the amplitude of the chaotic magnetic fields excited by the turbulence of neutral gas is higher than
the strength of large scale magnetic fields in the halo, the degree of polarization is small, around $3-4$\%.
If the kinetic energy of turbulent motions is relatively small, the large scale fields are not perturbed significantly by the turbulence.
In this case the upper limit of linear polarization of the synchrotron component may reach the level of 75\%.
\item  An alternative model for hydrogen ionization in Sgr B2 is irradiation by high energy electrons.
The energy of these electrons should be around or above 1 MeV.  These electrons generate a bremsstrahlung X-ray emission with
a spectrum $\propto E_x^{-1}$ which is harder than the emission from the Compton echo $\propto E_x^{-2}$ as observed by Suzaku and INTEGRAL.
Therefore, we expect that the bremsstrahlung component can be seen at high energies.
Our estimations show that the calculated bremsstrahlung component should be seen even in the 2009 data in the range 20-60 keV
if it is generated by electrons. If observations do not show any excess above the spectrum $E^{-2}$, then it is an argument against the
leptonic origin of hydrogen ionization at least in Sgr B2, although more detailed analysis of the X-ray data after 2009 is needed.
\end{itemize}

\section*{Acknowledgements}
%\acknowledgments
The authors would like to thank M. Goto, K. Mori, D. Prokhorov, T. Oka, D. Strelkov, and A. Strong for comments,
which helped us a lot to perform this analysis.
V.A.D. and D.O.C. acknowledge a partial support from the MOST-RFBR grant 15-52-52004 and the RFBR grant 15-02-02358.
D.O.C. and A.M.K. are supported in parts by the LPI Educational-Scientific Complex and Dynasty Foundation.
K.S.C. is supported by the G.R.F. Grants of the Government of the Hong Kong SAR under HKU 701013.
C.M.K. is supported in part by the Taiwan Ministry of Science and Technology Grants MOST 102-2112-M-008-019-MY3 and
MOST 104-2923-M-008-001-MY3.
C.Y.H. is supported by the National Research Foundation of Korea through grant 2014R1A1A2058590.
K.K.N. is supported by Research Fellowships of JSPS for Young Scientists.
V.A.D., D.O.C., A.M.K., K.S.C. and C.Y.H. acknowledge support from the International Space Science Institute-Beijing to the International Team
"New Approach to Active Processes in Central Regions of Galaxies".

\end{document}